\documentclass[12pt]{article}
\usepackage{amssymb} 
\begin{document}
\def\d{\delta}
\def\t{\theta}
\def\tbar{\bar{\theta}}
\def\del{\partial}
\def\qu{\mathcal{Q}}
\def\q{\mathit{Q}}
\def\l{\lambda}
\def\s{\sigma}
\def\sbar{\bar{\sigma}}
\def\w{\mathcal{W}}
\def\wbar{\bar{\mathcal{W}}}
\def\a{\alpha}
\def\b{\beta}
\def\g{\gamma}
\def\G{\Gamma}
\def\adot{\dot{\alpha}}
\def\bdot{\dot{\beta}}
\def\gdot{\dot{\gamma}}
\def\dbar{\bar{\delta}}
\def\kok{\sqrt{2}}
\def\yarim{{{1}\over{2}}}
%
%
\def\l{\lambda}
\def\p{\psi}
\def\f{\phi}
\def\vf{\varphi}
\def\ft{\phi^{\dag}}
\def\Ft{F^{\dag}}
\def\pbar{\bar{\psi}}
\def\lbar{\bar{\lambda}}
\def\sd{\mathcal{D}}
\def\se{\mathcal{E}}
\def\sf{\mathcal{F}}
\def\L{\mathfrak{L}}
\def\ds{\displaystyle}
\def\be{\begin{equation}}
\def\ee{\end{equation}}
\def\beq{\begin{eqnarray}}
\def\eeq{\end{eqnarray}}
\def\ov{\overline}
\thispagestyle{empty}
\begin{flushright}BN--TH--2000--09\\
\end{flushright}
\vskip5em
\begin{center}
{\Large{\bf On the Cohomological Structure of  Supersymmetric Lagrangeans With and Without Auxiliary Fields}}
\vskip1.5em
K. \"{U}lker\\\vskip3em
{\sl Physikalisches Institut der Universit\"at Bonn}\\
{\sl Nu{\ss}allee 12, D--53115 Bonn, Germany}\\
{\sl and}\\
{\sl Istanbul Technical University}\\
{\sl Science and Letters Faculty Physics Dept.}\\
{\sl 80626,  Maslak - Istanbul, Turkey}\\ 
\vskip1.5em
\end{center}

\vskip2em
%
%
\begin{abstract}

The origin of non-renormalization theorems in field theories with global supersymmetry can be traced to the fact that supersymmetric actions can be viewed as the highest components of respective supermultiplets. Supersymmetric interactions in particular can therefore be represented as supersymmetry variations of lower dimensional field polynomials. We investigate here this algebraic structure in the context of the Wess-Zumino model and N=1 and N=2 supersymmetric Yang-Mills theories.    
\end{abstract}
\vskip6em
{\small{e-mail:  ulker@th.physik.uni-bonn.de\\
\hspace*{1.3cm} kulker@itu.edu.tr}}
\vfill
\eject
\setcounter{page}{1}
%
\section{Introduction}

It is known that globally supersymmetric versions of renormalizable quantum field theories display important finiteness properties due to the cancellations of ultraviolet divergencies \cite{zum1}. These cancellations were first analyzed by Iliopuolos and Zumino \cite{iz} in the framework of supersymmetric Wess-Zumino models. Since then, with the development of supergraph techniques in superspace in which all particles in each supermultiplet are considered together as superfields \cite{fl,gsr,gs} , it became possible to display all simplifications due to supersymmetry.\\
\\On the other hand, one is forced to deal with the component field formalism in particular when calculations on non-trivial backgrounds are considered. However, to deal with supersymmetric theories in component picture several problems have to be overcome. First of all, in general the supersymmetry algebra realized without auxiliary fields will close only on-shell, that is modulo terms involving equation of motion of spinor fields. Second, the supersymmetric gauge theories in component gauges (i.e. Wess-Zumino gauge) display, besides this equation of motion terms, the additional characteristic complication that the algebra is modified by field dependent gauge transformations \cite{df}. The third problem is that the supergraph techniques give in many cases a manifest realization of non-renormalization theorems (i.e. the non-renormalization theorems of the chiral vertex in Wess-Zumino model), however these  non-renormalization theorems are by no means obvious in the component field formalism.\\
\\The first problem is cured \cite{ps,w,mag,mpw} by some sort of Batalin-Vilkovisky formalism \cite{bv} that is often called Algebraic Renormalization \cite{pigkit,sorlec}, whereas the second one is solved by extending BRS transformations to include supersymmetry transformations \cite{w,mag,mpw}. (For the problem of how to extend the BRS formalism to include arbitrary global symmetries, see \cite{bra1}.) Finally, to get a handle on the non-renormalization theorems in the component picture one has to note that the algebraic source of these theorems is to be found in the cohomology of supersymmetric theories: supersymmetric interaction terms being the highest component of some supermultiplet can be represented as a multiple supervariation of a lower dimensional field polynomial \cite{fk}.\\
\\In this note we study aforementioned algebraic structure in the context of Wess-Zumino model and N=1,2 supersymmetric Yang-Mills theories. The organization of the paper is as follows: in section 2 we discuss in which way the cohomological aspects are realized in Wess-Zumino model without auxiliary fields. In section 3 we generalize the results of \cite{fk} and section 2 to N=1 and N=2 pure Yang-Mills theories. In particular we show that the N=2 supersymmetric Yang-Mills Lagrangean can be represented as a fourfold supersymmetry variation of a gauge invariant field monomial canonical dimension two. It follows from this, according to an argument due to Bellisai et.al \cite{sor}, that the N=2 gauge coupling remains without renormalization beyond one loop order.    
\vfill
\eject
\section{The Wess Zumino Model}
We start to analyze the cohomological aspects of the non-renormalization theorems in Wess-Zumino (WZ) model with auxiliary fields first \cite{fk}. The Wess-Zumino model consists of the chiral multiplet \(\Phi=(\f,\, \p,\, F)\) and its complex conjugate, the anti-chiral multiplet \(\bar{\Phi}=(\f^{\dag},\, \bar{\p},\, F^{\dag})\). The action 
\footnote{Throughout the paper we use Wess-Bagger conventions where the metric is given by \(\eta_{\mu\nu}=(-1,1,1,1)\) \cite{WB}}   
\beq
S[x]&=& \int d^4 x (\L_{kin} + \L_{int})       \nonumber\\   &=&\int {d^4} x    \lbrace -\del_{\mu}\f^{\dag}\del^{\mu}\f                      - i\pbar  \bar{\del}\!\!\! / \p + F^{\dag} F  + [ m(\f F -{{1}\over{2}}\p\p)  +g(\f\f F - \p \p \f) + h.c.  ]                                                            \rbrace  
\eeq                           
is invariant under N=1 off-shell supersymmetry transformations:

\be
\d_{\a} \f = \kok \p_{\a} \qquad    \dbar_{\adot} \f= 0
\ee

\be
\d_{\a} \ft = 0 \qquad   \dbar_{\adot} \ft= \kok \pbar_{\adot}
\ee

\be
\d_{\a}\p^{\b}=\kok {\d_{\a}}^{\b}F  \qquad          \dbar^{\adot}\p^{\b}=-i\kok \sbar^{\adot\b}_{\mu}\del^{\mu}\f
\ee

\be
\d_{\a}\pbar_{\bdot}=-i\kok \s_{\a\bdot}^{\mu}\del_{\mu}\ft \qquad   \dbar^{\adot}\pbar_{\bdot}=\kok {\d^{\adot}}_{\bdot}\Ft
\ee

\be
\d_{\a} F= 0 \qquad       \dbar^{\adot}F = i\kok \sbar^{\adot\b}_{\mu}\del^{\mu} \p_{\b} 
\ee

\be
\d_{\a}\Ft= i\kok  \s_{\a\bdot}^{\mu}\del_{\mu}\pbar^{\bdot}\qquad                       \dbar^{\adot}\Ft = 0
\ee

\be
\{\d,\d\}\vf \, =\,0 \,= \, \{\dbar,\dbar\}\vf , \quad  \{\d_{\a}, \dbar_{\adot}\} \vf= -2i\s^{\mu}_{\a\adot}\del^{\mu}\vf
\ee
where \( \vf = \f, \p, F, \ft, \pbar, \Ft\).
The kinetic term, mass term and the interaction term in equation (1), being separately supersymmetric invariant, can be represented as supersymmetry variation of a lower dimensional field polynomial \footnote{For a similar approach of constructing N=1 globally and locally supersymmetic actions by BRS cohomological means and also for the discussion of anomalies, see \cite{bra2}}. We concentrate on the interaction part of the Lagrangean (1). 
\be
( \f\f F - \p \p \f)=-{{1} \over {12}} \d \d (\f\f\f) \qquad (\ft\ft \Ft - \pbar \pbar \ft )= -{{1} \over {12}} \dbar\dbar(\ft\ft\ft)
\ee
That is, the interaction terms are the highest components of a chiral and anti-chiral multiplet respectively of which the lowest components are \( \f^{3} \) and \(  \f^{\dag 3} \) respectively. The relations (9) were identified in ref.\cite{fk} as the algebraic source of the non-renormalization theorems in the Wess-Zumino model (with auxiliary fields). \\
The Wess-Zumino model is renormalizable. It is characterized by the supersymmetry Ward identities 
\be
\w_{\a} S= \int d^4 x{\d_{\a}\varphi{{\del}\over{\del\varphi}}}S=0 \qquad  \wbar_{\adot}S=      \int d^4 x {\dbar _{\adot} \varphi {{\del} \over{\del\varphi}} S } =0
\ee
that define the Green functions to all orders of perturbation theory where \(\G(\Phi, \bar \Phi)=S+\G_1+...\) denotes the generating functional of one particle irreducible (1PI) Green functions. Free parameters are fixed by  following normalization conditions\footnote{Here \( \G_{\vf_1...\vf_m}(x_1...x_n):={{\d^n}\over{\d\vf_1...\d\vf_n}}\G\big|_{all\,\vf=0}\) denotes the n-point 1PI Green functions.}:
\beq
\G_{F \Ft} (p^2) \Big|_{p^2 = \kappa^ 2} = 1 \nonumber\\\G_{F\f} (p^2) \Big|_{p^2 = 0} = m \\\G_{F\f\f} (p_1, p_2, p_3)\Big|_{p_i = 0}  =  - \frac \lambda 2   .\nonumber 
\eeq
The classical action of Wess-Zumino model is also invariant under a modified R'-symmetry \cite{fk}, besides the ordinary (conformal) one, which fixes the number of chiral and anti-chiral vertices for a given loop order and a given configuration of external fields.  One finds in this way for example that the L-loop contribution to the chiral vertex functional \(\G_{\f\p\p}\) contains L anti-chiral insertions: 
\be
\G_{\f\p\p}^{(L)}\big|_{g={\bar{g}}} \sim  \big( [\int d^4 x \bar{Q} ]_4 ^{(L)}.\G \big)_ {\f\p\p}  \sim ( [\int d^4 x (\ft\ft \Ft - \pbar \pbar \ft)]_4 ^{(L)}  \G \big)_ {\f\p\p}    .
\ee
Here the subscript 4 refers to the  Zimmerman subtraction prescription for operators of canonical dimension 4. Now with the help of eq.(9) and (12) we obtain 
\be
\G_{\f\p\p} ^{(L)} \sim ( [\int d^4 x \dbar \dbar (\f^{\dag 3} )][...] ^{(L-1)} \G \big)_ {\f\p\p}
\sim [\int d^4 x \dbar (\ft\ft\ft)][...] ^{(L-1)}\G ) _{\dbar(\f\p\p)}
\ee
where \(\dbar(\f\p\p)\sim \f\del_{\mu}\f\p \). For the second proportionality in (13) we use the supersymmetric Ward identity. It was shown in \cite{fk} that these formal manipulations are compatible with algebraic renormalization.  One derives from (13) two consequences: 1.) \(\G_{\p\p\f}\) is a superficially convergent vertex function since on the r.h.s. of (13) appears  an insertion of canonical dimension \({{7}\over{2}}\) and the derivative on the external scalar field does not contribute to the superficial degree of divergence. 2.) The vertex function vanishes at zero momentum.\\
\\It may be instructive to discuss the non-renormalization theorem in a formalism where the auxiliary fields  \(F \) and \(\Ft \) are eliminated by their equation of motion.
\be
F = -m\ft - {\bar g} \ft\ft \qquad \Ft = -m\f -g\f\f 
\ee
Inserting these expressions into the action (1) one arrives at 
\be
S= \int d^4 x [-\del_{\mu}\ft \del^{\mu}\f   -i\pbar \del\!\!\! / \p                                           -\yarim m\p\p     -\yarim \pbar \pbar                                                               -g\p\p\f -\bar{g}\pbar\pbar\ft   -( m\ft +{\bar{g}} \ft\ft )( m\f +g\f\f)]
\ee
One may try again to represent interactions as supersymmetry variations of lower dimensional terms. For this purpose we have to eliminate the auxiliary fields above the equations (2-8). Then the modifications read as 

\be
\d_{\a} \p ^{\b}= \kok \d_{\a}^{\b}(-m\ft -\bar{g} \ft\ft) \quad \d_{\adot} \p ^{\bdot}=\kok \dbar_{\adot}^{\bdot}  ( -m\f -g\f\f)    .
\ee
It appears that the interaction part of the action (15) can no longer be decomposed into invariant chiral and anti-chiral parts. And a fortiori the cohomological identities (9) have no direct analogue. That is the terms \(\d\d \f ^3 \) and \( \dbar \dbar \f ^{\dag 3} \) are no longer supersymmetric invariant since \(\dbar\) applied to  \(\d\d\f^3\) ( or \(\d \) applied to \( \dbar \dbar \f ^{\dag 3} \) ) produces besides a derivative term an additional disturbing equation of motion term, as the supersymmetry algebra now closes only on-shell. But we have instead the identities 
\be
-{{1}\over {12}} \d\d\f^3={{d}\over{dg}}\L_{int}\qquad   -{{1}\over {12}} \dbar \dbar \f ^{\dag 3} = {{d}\over{d\bar{{g}}}} \L_{int}    .
\ee
This allows to write 
\be
{{d}\over{d\bar{{g}}}} \G_{\f\p\p} \sim [\int d^ x \dbar \dbar    \f ^{\dag 3} ] \G_{\f\p\p} \sim -[\int d^ x \dbar \f^{\dag 3}] \G_{\dbar(\f\p\p)}
\ee
and an analogous chain of equations for the anti-chiral vertex functions. We conclude as before that the derivatives of the vertex functions are superficially convergent. The ensuing integrations will not affect this conclusion.\\ 
\\Recent explicit calculations of N. vonRummel \cite{vr} in the component picture confirm the non-renormalization theorems. But this author claims that the chiral vertex functions do not vanish at zero momentum\footnote{We thank N. Von Rummel for explaining his results.}. This would mean that the treatment in the component picture induces a finite shift in the renormalized parameters of the model in comparison to the off-shell approach. It seems worthwhile to clarify this issue.    
%
%
\section{Super Yang-Mills Theories}
\subsection{N=1 Pure Yang-Mills Theory}
	The action of the pure super Yang-Mills theory contains the N=1 vector multiplet \((A_{\mu} ,\l , \lbar ,\sd )\) and reads\footnote{We take, for simplicity,  SU(2) gauge group and use the following SU(2) matrix notation throughout this work: 
\[  \Phi = \Phi^{a} \tau^{a}, \quad [ \tau^{a}, \tau^{b} ]= i f^{abc} \tau^{c}, \quad Tr( \tau^{a} \tau^{b} ) = \yarim \d^{ab} \]  \[D_{\mu}{\Phi}=\partial_{\mu}{\Phi}-i[A_{\mu},\Phi] \] \[F_{\mu\nu}=\partial_{\mu}A_{\nu}-\partial_{\nu}A_{\mu}-i[A_{\mu},A_{\nu}] \] where $f^{abc}$ is the totally antisymmetric matrix. },
\be
S_{YM}= Tr\int d^4 x(-\yarim F_ {\mu\nu} F^ {\mu\nu}-2i{\l}D\!\!\!\! / \bar{\l} +\sd ^2 )   
\ee
The SUSY transformations for N=1 vector multiplet is given by,
\be 
\q A_{\mu}=i{\t}{\s} _{\mu} \bar{{\l}}+ i\bar{\t}\bar{\s} _{\mu}{\l}
\ee

\be
\q{\l}={\s}^{\mu\nu}{\t}F_{\mu\nu}+ i\t \sd
\ee

\be
\q{\lbar}={\sbar}^{\mu\nu}{\tbar}F_{\mu\nu}- i\tbar \sd
\ee

\be
\q\sd= -\t D\!\!\!\!/\bar{\l} +\tbar \bar{D}\!\!\!\!/ \l
\ee
where
\be
\q = \qu + \bar{\qu} = \t^\a \d_{\a} + \tbar_{ \adot} \dbar^{ \adot}
\ee
and \(\t, \tbar \) are supersymmetry parameters. The algebra reads
\be 
\{ \d,\d \} \vf \, =\, 0 \,=\, \{ \dbar ,\dbar \} \vf \quad \{ \d,\dbar \} \vf \,=\, -2i\s^{\mu} D_{\mu} \vf      . 
\ee
By analyzing supersymmetry transformations of the fields one can see that the action can be found from the gauge invariant, dimension 3 chiral and anti-chiral field polynomials \( \l^{\a} \l_{\a}\) and \( \lbar_{\adot} \lbar^{\adot} \)  : 

\be
S_{YM} = {{1}\over {8}} Tr \int d^4 x \{ (\d \d ) (\l \l ) + (\dbar \dbar ) (\lbar \lbar ) \}   . 
\ee
The auxiliary fields \(\sd\) are crucial for the latter irrespective of the fact that they vanish on-shell. One finds then instead \({{1}\over {8}} Tr ( (\d \d ) (\l \l ) + (\dbar \dbar ) (\lbar \lbar ) ) \) is equal to 

\be
Tr [ -\yarim F_{\mu\nu} F^{\mu\nu} -{{3i}\over{4}} \l D\!\!\! / \lbar -{{3i}\over{4}} \lbar \bar{D\!\!\! /} \l ]     . 
\ee
It seems advisable to stick to the off-shell formulation. It is in our opinion still a challenge to derive (or redrive) the Shiffman-Vainshtein formula \cite{shif} for the \(\b\) function of the gauge coupling.     
\subsection{N=2 Pure Yang-Mills}

It is well known that twisting of the ordinary N=2 super Yang-Mills (SYM) theory, in the Wess-Zumino gauge, yields  topological field theories of the cohomological type \cite{wit} which have zero degrees of freedom and whose correlators are independent of the metric of the manifold where the theory is defined. On the other hand it is found out that ordinary supersymmetric gauge theories have also a class of position independent correlation functions \cite{am}. Indeed twisting procedure is nothing else than a variable redefinition and one can realize that the correlation functions of the topological theory coincide with the subset of those of the physical theory. Recently this relation was used to study the non-renormalization of the N=2 SYM theory in the twisted formulation \cite{sor} which is our motivation in this section.\\
\\The action of the pure super Yang-Mills contains the N=2 \cite{soh} vector multiplet \[(\f ,\ft, A_{\mu}, \p, \pbar , \l , \lbar, \sd, \se, \sf ) \] where all the components of it transform under adjoint representation of the gauge group.The action with a topological term is given by, 

\beq
S&=&Tr \int d^4 x [-\yarim F_{\mu\nu} F^{\mu\nu} -{{i}\over {4}} \epsilon^{\mu\nu\kappa\l} F_{\mu\nu}F_{\kappa\l} -2i \l D \!\!\!\! / \lbar -2i\p D \!\!\!\!/ \bar{\p} +2\f D_{\mu} D^{\mu} \f^{\dag} \nonumber\\ && +2i \kok (\p [\l , \ft ] -\bar{\l}  [ \bar{\p}, \f ] ) -[\f , \f^{\dag} ]^2 +\sd ^2 +\se ^2 +\sf ^2]
\eeq
and N=2 off-shell SUSY transformations \cite{soh} read in Wess-Bagger \cite{WB} conventions as follows: 
\beq
\q A_{\mu}=    i{\t_1}{\s} _{\mu} \bar{{\l}}       + i{\t_2}{\s} _{\mu} \bar{{\p}}          +  i\bar{\t_1}\bar{\s} _{\mu}{\l}            +    i\bar{\t_2}\bar{\s} _{\mu}{\p}
\eeq

\beq
\q {\l}=      {\s}^{\mu\nu} {\t_1} F_{\mu\nu}        + i{\t_1} [{\f},{\f}^\dag]             -i\sqrt{2}{\s}^{\mu}\bar{\t_2}D_{\mu}{\f} 
                  +i{\t_2}\sd+{\t_2}\se+i{\t_1}\mathcal{F}
\eeq   

\beq
\q \p=    \s^{\mu\nu} {\t_2} F_{\mu\nu}        + i\t_2[\f,\f^{\dag}]           + i\sqrt{2}\s^{\mu} {\bar{\t_1}} D_{\mu} \f              + i{\t_1} \mathcal{D} - {\t_1}\mathcal{E}                    -  i{\t_2}\mathcal{F}
\eeq

\beq
\q \f= \sqrt{2}\t_1\p  -   \sqrt{2}\t_2\l
\eeq

\beq
\q \mathcal{D}=       -\t_1\s^{\mu}D_{\mu}\bar{\p}            - \t_2\s^{\mu}D_{\mu}\bar{\l}             + \sqrt{2}\t_1[\l,\f^{\dag}]                  - \sqrt{2}\t_2[\p,\f^{\dag}]\nonumber\\                  + \bar{\t_1}\bar{\s}^{\mu}D_{\mu}\p                       +  \bar{\t_2}\bar{\s}^{\mu}D_{\mu}\l                      -  \sqrt{2}{\bar{\t_1}} [\bar{\l},\f]                         +     \sqrt{2} {\bar{\t_2}} [\bar{\p},\f]
\eeq

\beq
\q \mathcal{E}&=&       -i\t_1\s^{\mu}D_{\mu}\bar{\p}+                i\t_2\s^{\mu}D_{\mu}\bar{\l}+                  i\sqrt{2}\t_1[\l,\f^{\dag}]+                    i\sqrt{2}\t_2[\p,\f^{\dag}]\nonumber\\                   &&- i\bar{\t_1}\bar{\s}^{\mu}D_{\mu}\p   +                        i\bar{\t_2}\bar{\s}^{\mu}D_{\mu}\l                           + i\kok \bar{\t_1} [\bar{\l} , \f ]   +                               i\sqrt{2} \bar{\t_2} [\bar{\p},\f ]
\eeq

\beq
\q \mathcal{F}=       -\t_1\s^{\mu}D_{\mu}\bar{\l}+         \t_2\s^{\mu}D_{\mu}\bar{\p}-   \sqrt{2}\t_1[\p,\f^{\dag}]- \sqrt{2}\t_2[\l,\f^{\dag}] \nonumber\\     + \bar{\t_1}\bar{\s}^{\mu}D_{\mu}\l-   \bar{\t_2}\bar{\s}^{\mu}D_{\mu}\p  + \sqrt{2} \tbar_1 [\bar{\p},\f] + \sqrt{2} \bar{\t_2} [\bar{\l},\f]
\eeq
here \(\mathcal{D}, \mathcal{E}, \mathcal{F}\) are the auxiliary fields and \(\q\) denotes the generators of supersymmetry transformations,
\beq
\q &=& \qu_1 + \qu_2 + \bar{\qu_1}+ \bar{\qu_2}\nonumber\\ &=& \t_1^\a \d_{1\a} +  \t_2 ^\a \d_{2\a} +      \tbar_1 { \adot} \dbar_1 ^{ \adot} + \tbar_2 { \adot} \dbar_2 ^{ \adot}   .
\eeq
They satisfy the following algebra,

\be
[\qu_i ,\bar{\qu_j} ] = -2i(\t_i \s^{\mu} \bar{\t_j} )D_{\mu}
\ee

\be
[\qu_i ,\qu_j] A_{\mu} = -2\kok \varepsilon_{ij} (\t_i \t_j ) D_{\mu} \ft 
\ee

\be
[\bar{\qu_i} ,\bar{\qu_j}] A_{\mu} = -2\kok \varepsilon_{ij} (\tbar_i \tbar_j ) D_{\mu} \f 
\ee

\be
[\qu_i,\qu_j]\,  \Xi = -2i\kok \varepsilon_{ij} (\t_i \t_j ) [\Xi ,\ft ] 
\ee

\be
[\bar{\qu_i},\bar{\qu_j}]\,  \Xi = -2i\kok \varepsilon_{ij} (\t_i \t_j ) [\Xi ,\f ] 
\ee

\[ i,j\, = \, 1,2  \quad \varepsilon_{12} \, = \, -\varepsilon_{21}\, =\, 1\] 
\[\Xi=\f ,\ft ,\l ,\lbar ,\p ,\pbar ,\sf ,(\se +i\sd) ,(\se -i\sd) \]
Equation (37) is gauge covariant extension of the usual supersymmetry algebra whereas the r.h.s of (38-41) may be interpreted as central extensions in the form of gauge transformations with the gauge parameters \(\f\) and \(\ft\) respectively.\\
\\The renormalization of the N=2 model has been discussed by Maggiore \cite{mag} using above mentioned technique of an extended BRS formalism \cite{w}.The topological, twisted version of the model has been treated in a similar spirit most recently by Blasi et.al. \cite{sor} by relating the gauge invariant polynomial \( Tr(\f ^2) \) to the twisted N=2 Yang-Mills action \cite{sorlec}. Their providing for non-renormalization of the coupling constant beyond one-loop can easily be transported in the ordinary (untwisted) case. The unique reason for the present non-renormalization theorem is to be found in the fact that the N=2 Yang-Mills action can be written as an integral of a forth super variation of dimension two filed monomial \footnote{ To get rid of the topological term one simply adds the hermitean conjugate of eq.(42) since it gives the same action that has the opposite sign for the topological term.}: 
\be
S_{N=2}= {{1}\over {16}} Tr \int d^4 x (\d_1 \d_1 )(\d_2 \d_2 ) Tr (\f^2 ) 
\ee
that is the highest component of a N=2 chiral multiplet \cite{soh} of which the highest component is \( Tr(\f ^2) \) .\\
\\To find this representation it is crucial, as before in the N=1 YM theory, that an auxiliary field formalism is available. The representation in (42) can be verified by noting that this expression has canonical dimension four and it is gauge invariant and supersymmetric. There is only one such expression which is N=2 SYM action.\\
\\On the other hand by using on-shell transformations, i.e. by setting auxiliary fields \(\sd ,\se ,\sf \) to zero, a similar fourfold supersymmetry variation of \( Tr \f ^2 \) yields \footnote{One has to distinguish the different combinations of \(\d_1   \) and \( \d_2 \) due to the fact that relations (38-41) hold only on-shell since the r.h.s of these relations become gauge transformations plus the equation of motion of the fermions.}
\beq
{{1} \over {96}}Tr \int d^4 x ( {\d_1}^\a  {\d_1}_\a {\d_2}^\b  {\d_2}_\b - {\d_1}^\a  {\d_2}^\b {\d_1}_\a  {\d_2}_\b + {\d_1}^\a  {\d_2}^\b {\d_2}_\b  {\d_1}_\a  \quad + \quad (1 \longleftrightarrow 2)) \f^2 \nonumber\\                                                                      =Tr \int d^4 x (-\yarim F_{\mu\nu} F^{\mu\nu} -{{i}\over {4}} \epsilon^{\mu\nu\kappa\l} F_{\mu\nu}F_{\kappa\l} -i \l D \!\!\!\! / \lbar - i\p D \!\!\!\!/ \bar{\p} + \f D_{\mu} D^{\mu} \f^{\dag} -i \kok \bar{\l}  [ \bar{\p}, \f ]  )  . 
\eeq
Note that the difference between the action (28) and the expression (43) can be written in terms of equations of motion of \(\l ,\p \) and \(\f \) as following \footnote{ A similar argument also holds for Wess-Zumino model and N=1 super Yang-Mills theory when one tries to construct the corresponding actions by using on-shell transformations.} :
\beq
Tr \int d^4 x \{ -i \p \, (D \!\!\!\!/ \bar{\p} - \kok [\l, \f^{\dag}]) -i \l \, ( D \!\!\!\! / \lbar + \kok [\p , \f^{\dag}]) + \f \, (D_{\mu} D^{\mu} \f^{\dag} -i \kok [\bar{\l}  ,\bar{\p}] - [ \f^{\dag}, [\f , \f^{\dag}]] )  \} .\nonumber
\eeq
Finally, it is worthwhile to mention that after twisting the expression (43) (by identifying the susy index with the undotted spinor index), one gets the Topological Yang-Mills action in the approach of Baulieu-Singer \cite{bs} for a particular choice of Faddeev-Popov ghosts since the coefficient of \( F_{\mu \nu} F^{\mu \nu} \) term is arbitrary in this approach.\\
\\The completely open problem that remains is the discussion of the non-renormalization theorems of N=2 super Yang-Mills theories with matter hypermultiplets and N=4 supersymmetric models due to the fact that there are no known auxiliary field formulation of these models. We hope to disentangle this fact in a future publication.\\
\\{\bf Acknowledgements: }
The author is grateful to R. Flume for suggesting this work and sharing his valuable knowledge while this work was in progress. Special thanks are due to M. Hortacsu for his continuous interest and also for many discussions.\\This work is supported by TUBITAK, Scientific and Technical Research Council of Turkey under BAYG-BDP. 
 
 
\end{document}